\documentclass[useAMS,usenatbib,aps,floatdix,nofootinbib]{mn2e}

\usepackage{graphics}
\usepackage{amssymb,amsmath}
\usepackage{amsmath}
\usepackage{psfrag}
\usepackage{graphicx}
\usepackage{epsfig}
\usepackage{epstopdf}

\usepackage{color}
\usepackage{ulem}

\newcommand{\f}{\frac}

\newcommand{\be}{\begin{equation}}      
\newcommand{\ee}{\end{equation}}      
      
\newcommand{\bef}{\begin{figure}}      
\newcommand{\eef}{\end{figure}}      
\newcommand{\bea}{\begin{eqnarray}}    
\newcommand{\eea}{\end{eqnarray}}      
  
\def\spose#1{\hbox to 0pt{#1\hss}}
\def\ltapprox{\mathrel{\spose{\lower 3pt\hbox{$\mathchar"218$}}
\raise 2.0pt\hbox{$\mathchar"13C$}}}
\def\gtapprox{\mathrel{\spose{\lower 3pt\hbox{$\mathchar"218$}}
\raise 2.0pt\hbox{$\mathchar"13E$}}}
\def\inapprox{\mathrel{\spose{\lower 3pt\hbox{$\mathchar"218$}}
\raise 2.0pt\hbox{$\mathchar"232$}}}

\def\bse{\begin{subequations}}
\def\ese{\end{subequations}}

\def\lsim{\raise 0.4ex\hbox{$<$}\kern -0.8em\lower 0.62ex\hbox{$\sim$}} 
\def\gsim{\raise 0.4ex\hbox{$>$}\kern -0.7em\lower 0.62ex\hbox{$\sim$}}

\def\f0N{f_0^{(N)}}
\def\bec{\begin{center}}
\def\eec{\end{center}}

\title[Non-linear clustering: 1D toy models]
{Non-linear gravitational clustering of cold matter in an expanding universe: indications from 1D toy models}
\author[M. Joyce and F. Sicard]
{Michael Joyce${^{1,}}{^2}$ and Fran\c{c}ois Sicard${^{1}}$\\
$^1$Laboratoire de Physique Nucl\'eaire et Hautes \'Energies,
Universit\'e Pierre et Marie Curie - Paris 6, CNRS IN2P3 UMR 7585, \\  4 Place Jussieu, 75752 Paris Cedex 05, France\\
$^{2}$Laboratoire de Physique Th\'eorique de la Mati\`ere Condens\'ee,
Universit\'e Pierre et Marie Curie - Paris 6,
CNRS UMR 7600, \\ 4 Place Jussieu, 75752 Paris Cedex 05, France
}

\begin{document}

\date{\today}

\maketitle

\begin{abstract}
Studies of a class of infinite one dimensional self-gravitating systems have highlighted 
that, on the one hand, the spatial clustering which develops may have scale invariant (fractal) 
properties, and, on the other, that they display ``self-similar" properties in their temporal evolution.
The relevance of these results to three dimensional cosmological simulations
has remained unclear. We show here that the measured exponents 
characterizing the scale-invariant non-linear clustering are in excellent agreement
with those derived from an appropriately generalized ``stable-clustering" hypothesis.
Further an analysis in terms of ``halos"  selected  with a friend-of-friend algorithm
reveals that such structures are, statistically, virialized across the range of
scales corresponding to scale-invariance. Thus the strongly non-linear clustering in 
these models is accurately described as a virialized fractal structure, very much
in line with the ``clustering hierarchy"  which Peebles originally envisaged 
qualitatively as associated with stable clustering. If transposed 
to three dimensions these results would imply, notably, that cold dark  matter halos 
(or even subhalos)  are 1) not well modeled as smooth objects, and 2) that the 
supposed ``universality" of their profiles is, like apparent smoothness, an artefact 
of poor numerical resolution.
\end{abstract}


\section{introduction}


The much acclaimed 
successes of the $\Lambda$CDM cosmology in matching many observations
concern essentially its homogeneous limit and the linear regime in which
perturbations to homogeneity are small. The success of the model in accounting
for the numerous observations which probe the non-linear regime, where density
fluctuations are large, is much more uncertain.
An important consideration in this respect is the great difficulty of calculating the 
model's predictions in this regime. Even in the idealized limit in which clustering 
arises from gravity alone, predictions reside solely on numerical simulations. 
The latter have, despite impressive increases in their size and sophistication, 
still quite limited spatial resolution. Further, essentially because there are no
non-trivial analytical benchmarks which they can be tested against, it remains
unclear to what extent  their results are conditioned by
these resolution limits.
In this paper  we explore what might be learnt about the nature of non-linear 
gravitational clustering from the study of a class of one dimensional (1D) models, 
which have the interest of offering, even with modest computer resources,
very much greater spatial resolution and exact numerical integration.
An early study in this spirit is that of  \cite{melott_prl_1982,melott_1d_1983}  
which used such a 1D model to explore clustering in hot dark matter cosmologies.
 \cite{rouet_etal} derived and studied a slightly different model
to the most naive 1D version of clustering in an 
Einstein de Sitter (EdS) universe considered 
by \cite{yano+gouda, aurell+fanelli_2002a, aurell_etal_2003}.
\cite{valageasOSC_2}  and \cite{agmjfs_pre2009} have studied the 
version without expansion, while Miller et al. 
\citep{miller+rouet_2002, miller+rouet_2006, miller_etal_2007, miller+rouet_2010a}
have reported results on all three of these models.
Despite these studies it remains unclear, however,  whether a good, and really 
useful, analogy can be made with 3D cosmological simulations. Generalizing 
the results of \cite{ yano+gouda} and \cite{agmjfs_pre2009}, we confirm here
the very strong qualitative similarities in the temporal development of 
clustering to that in the 3D case. Further we show the scale invariant
properties of the clustering in real space,  emphasized and 
analyzed by \cite{ miller+rouet_2002, miller+rouet_2006,  miller_etal_2007}, 
can in fact be well explained within an analytical framework
similar to one proposed long ago for the 3D case \citep{peebles_1974, davis+peebles_1977, peebles}.
We derive the appropriate 1D version of this 
generalized ``stable clustering" prediction for the correlation exponents, and 
show it to provide a very good approximation to the numerical results. 
We also show that, in a precise sense, the strongly non-linear
clustering can be properly described as a {\it virialized} fractal hierarchy, 
in line with what was originally anticipated qualitatively in three dimensions. 
We argue finally for the extrapolation of these conclusions about
the qualitative nature of clustering to the relevant 3D case, and 
consider its implications, in particular for what concerns the nature and 
properties of ``halos" in cold dark matter cosmologies. 

\section {1D versions of cosmological N-body simulations}  
Dissipationless cosmological  N-body simulations (for a review
see, e.g., \cite{bagla_review}) solve numerically the equations 
\begin{equation}
\label{3d-equations}
\frac{d^2 {\bf x}_i}{dt^2} +
2H \frac{d{\bf x}_i}{dt}  = -\frac{Gm}{a^3}{\sum}^J
\frac{{\bf x}_i - {\bf x}_j}{\vert {\bf x}_i - {\bf x}_j \vert^3} \,,
\end{equation}
where ${\bf x}_i$ are the comoving particle coordinates, $a(t)$ is the 
appropriate scale factor for the cosmology considered, with 
Hubble constant $H(t)={\dot a}/{a}$. The superscript `J'
in the sum, which runs in practice over the infinite system
constituted by periodic copies of a cube containing $N$
particles,  indicates that the (badly defined) contribution of
the mean mass density (which sources the Hubble expansion) 
is subtracted.  We consider here  simply the 1D system obtained by 
replacing the 3D Newtonian force term by the analogous 1D expression, derived
starting from the 1D Poisson equation which gives 
a force between particles in one dimension (or, equivalently, infinite 
parallel sheets embedded in three dimensions) independent 
of separation.
We thus consider the equations 
\begin{equation}
\label{1d-equation}
\frac{d^2 x_i}{dt^2} +
2H \frac{d x_i}{dt}= -\frac{g}{a^3}
\lim_{\mu \rightarrow 0} \sum_{j\neq i}
\textrm{sgn}(x_i - x_j) e^{-\mu \vert { x}_i - {x}_j \vert}, 
\end{equation}
where the limiting procedure used in the sum is simply a 
convenient way to explicit the subtraction of the background \citep{kiessling},
and $g$ is the coupling constant (with $g \equiv 2\pi \Sigma G$  
for sheets of surface mass density $\Sigma$). Initial conditions are 
generated, just as in 3D cosmological simulations, by applying 
appropriate small perturbations to particles initially on an infinite perfect 
lattice. If the displacement  from its initial lattice site of particle $i$ is 
$u_i$ it can be shown rigorously  \citep{agmjfs_pre2009} that 
the net force on the particle, until it crosses another one,  is exactly 
proportional to $u_i -\langle u \rangle$, 
where $\langle u \rangle$ is the average displacement. 
The equations of motion up to the time at which particles cross 
become
\begin{equation}
\label{1d-equation-u}
\ddot{u}_i +
2H \dot{u}_i= \frac{2gn_0}{a^3}u_i
\end{equation}
where $n_0$ is the mean particle density (and we take $\langle u \rangle=0$).
When the {\it  particles} in our model cross one another the force changes simply
by $\pm 2g$. However, since in 1D a crossing of particles is 
equivalent, up to exchange of labels, to an elastic ``collision" between them, one 
can instead consider (as we are not interested in the labels of the particles)
a system of particles which bounce elastically and follow Eq ~(\ref{1d-equation-u}) 
{\it at all times} between ``collisions". Finally, by a transformation of the time coordinate to
$\tau= \int dt/a^{3/2}$,  and appropriate choice of units of $\tau$,  
we can rewrite these equations as 
\begin{equation}
\label{1d-equation-u-final}
\frac{d^2 {u}_i}{d\tau^2} +
\Gamma \frac{d { u}_i}{d\tau} = u_i \,,
\end{equation}
i.e., as those of a set of damped inverted harmonic oscillators.  
We note that these equations coincide simply with those for displacements 
of {\it fluid} elements in the Zeldovich approximation (see, e.g. \cite{buchert2}),
which is in fact exact up to ``shell crossing" for 1D perturbations.
The same system can thus be derived as a simple analytical continuation 
of this approximation at shell crossing \citep{yano+gouda, aurell_etal_2003}.

For a generic cosmology $\Gamma$ in Eq.~(\ref{1d-equation-u-final})
is  a non-trivial function of $\tau$, but in the specific case of 
an EdS cosmology (for which $a \propto t^{2/3}$)  it is a constant, 
$\Gamma=1/\sqrt{6}$ (and $\tau \propto \ln a$).  The model 
derived and studied by  \cite{rouet_etal} corresponds instead to
the case $\Gamma=1/\sqrt{2}$, while the case $\Gamma=0$ 
(i.e. static limit) has also been studied by 
several authors (see references given above).  We note that,
despite the fact that we obtained the value $\Gamma=1/\sqrt{6}$
above, this is not necessarily {\it the} more appropriate
value to consider for our study:  in the derivation just given
the 3D Hubble law has been imposed {\it by hand}. 
This has been done because  the 1D Hubble law (which would
arise if one follows fully the analogy with the 3D derivation 
starting from  physical coordinates) gives a  completely different 
qualitative  behavior --- collapse in a finite time --- which has no 
relevance to 3D cosmology. Thus  in order to ``mimic" EdS 
cosmology, one could equally take the functional form of the 
corresponding Hubble law, but leave its normalization as a free 
parameter. This leads to the Eqs.~(\ref{1d-equation-u-final}) 
with $\Gamma$ an undetermined constant. These equations 
then simply define a simple class of toy  models, which may 
be useful for {\it qualitative} understanding of the 3D problem. 

Given initial displacements and velocities, Eqs.~(\ref{1d-equation-u-final})
can trivially be integrated exactly between particle collisions  
when $\Gamma$ is a constant. To determine the time of the
next collision (and which pair of particles it involves) requires
then only the solution of algebraic equations. The numerical
integration can therefore be performed ``exactly", i.e., up to
machine precision. As described by \cite{noullez_etal} the integration 
can be sped up optimally using a ``heap" structure. All the results 
presented here are for systems with $N=10^5$ particles (and
periodic boundary conditions).

In making the analogy with 3D simulations, which aim to reproduce
the {\it collisionless} limit  of gravitational clustering and include a
smoothing of the singularity in the 3D Newtonian force for this 
reason, it is to be noted that, on the time scales we will 
consider, these 1D systems, without force smoothing, can be 
expected to represent extremely well this limit. 
Studies of collisional relaxation in finite 1D self-gravitating 
systems (see  \cite{mjtw_relaxation_2010}, and references therein) have 
shown that it is very suppressed compared to that in 3D systems, a typical 
relaxation time for a virialized system of number density $n$ 
being $\sim (10^4 - 10^6) N/\sqrt{gn}$. Using this estimate 
in the simulations reported below, it is simple to show that 
even the relaxation time for the smallest and densest 
clusters is much longer than the duration of the simulation.
The reason for this relative suppression of this relaxation is 
that, in one dimension, there is no analogy to 3D 2-body 
relaxation:  the ``collisions" of particles we have discussed 
above do not contribute to the usual collision term in the 
Boltzmann equation. Indeed it is for this reason that
particle crossings and ``collisions" are equivalent.

\section{ Comparison with simulations in 3D: self-similarity}
Various studies (see references above) of these models starting from
close to uniform initial conditions have noted the striking qualitative
similarity of the clustering to that in three dimensions: one observes
the formation of structures (i.e. overdense regions) first at small scales
and then at ever larger scales. Let us consider first more closely the
temporal character of this evolution.
It is canonical and instructive in the study of cold dark matter
clustering in the universe to consider evolution from initial conditions 
characterized by power spectrum (PS)  $P(k) \sim k^n$ (with $n$ a constant), 
and velocities prescribed by the growing mode of linear perturbation 
theory. Although the ``real"  cosmological  power spectrum 
(in currently favoured {\it cold} dark matter models) is not 
exactly power-law it may be well approximated as such, with a 
slowly varying exponent, between $n=-1$ and $n=-3$
over the relevant range (see, e.g. \cite{peebles_1993}).  
The simulation of such ``scale-free" initial conditions is expected 
to give rise to simple behaviours (which one might hope to understand
analytically). In particular, for $-d<n<4$ (where $d$
is the spatial dimension) one expects theoretically \citep{peebles}
a  process of hierachical structure formation which 
\begin{itemize}
\item is driven by linear amplification of the initial fluctuations. For
power law initial conditions and the models we are considering, it is 
simple to show that this corresponds to the scale of non-linearity 
(at which perturbations are of order one) which grows in
proportion to
\begin{equation}
\label{RsubS-prediction}
R_s(t) = \exp \left( \frac{2 \lambda_{+} (\Gamma)} {1+n} \tau \right) \,,
\end{equation}
where $\lambda_+ {(\Gamma)= -\frac{\Gamma}{2} + \sqrt{\big(\frac{\Gamma}{2}\big)^2 +1}}$ 
is the root of the characteristic 
equation of Eq.~(\ref{1d-equation-u}) associated to 
the growing mode of linear theory;  
\item tends asymptotically to ``self-similar" behaviour in the 
whole non-linear regime, i.e., the temporal evolution of
all correlation properties can be obtained by a rescaling
of the spatial coordinates, in proportion to the same function
$R_s(t)$. 
\end{itemize}
\begin{center}
 \begin{figure}
  \includegraphics[angle=-90, width=1.0\columnwidth]{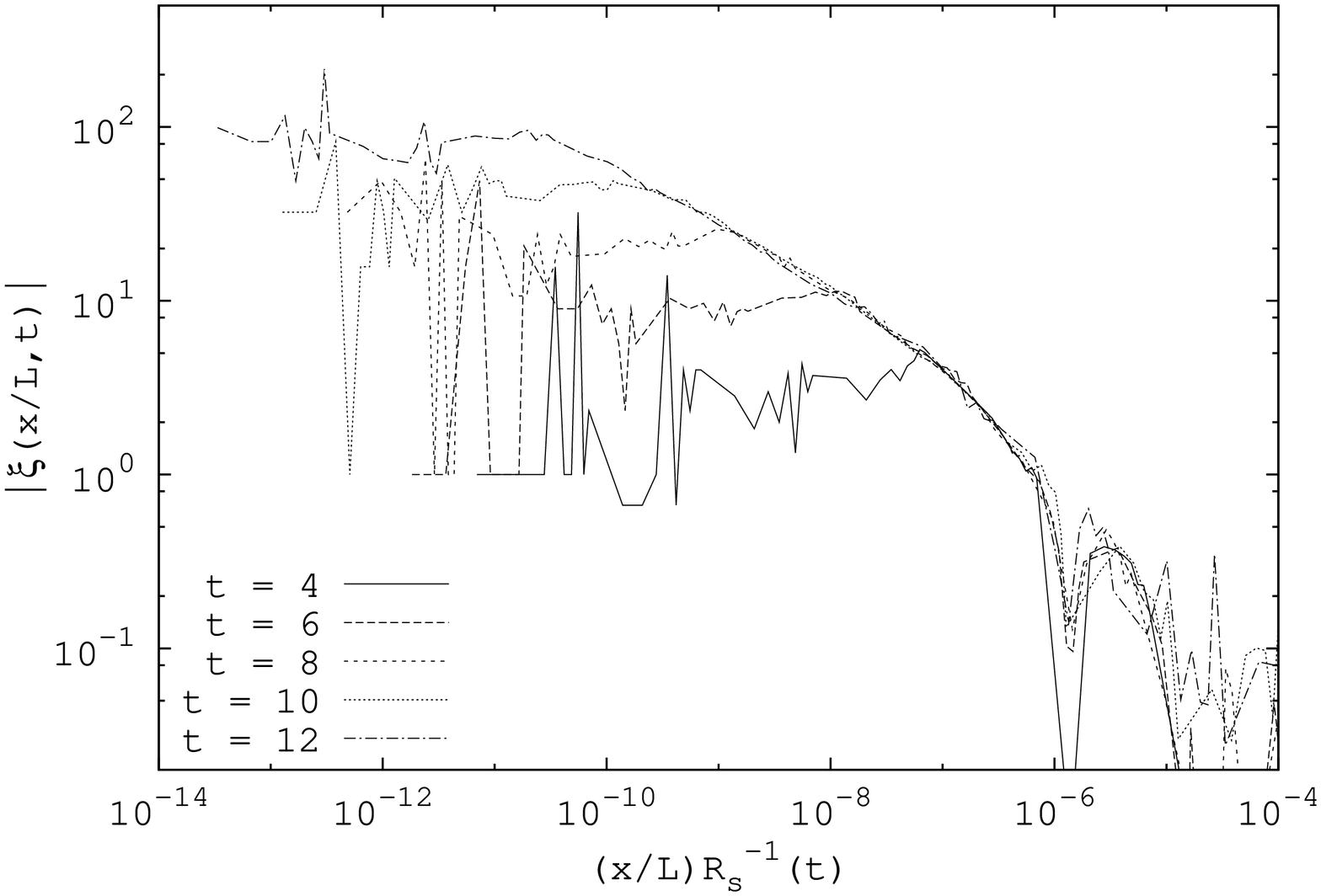}
  \includegraphics[angle=-90, width=1.0\columnwidth]{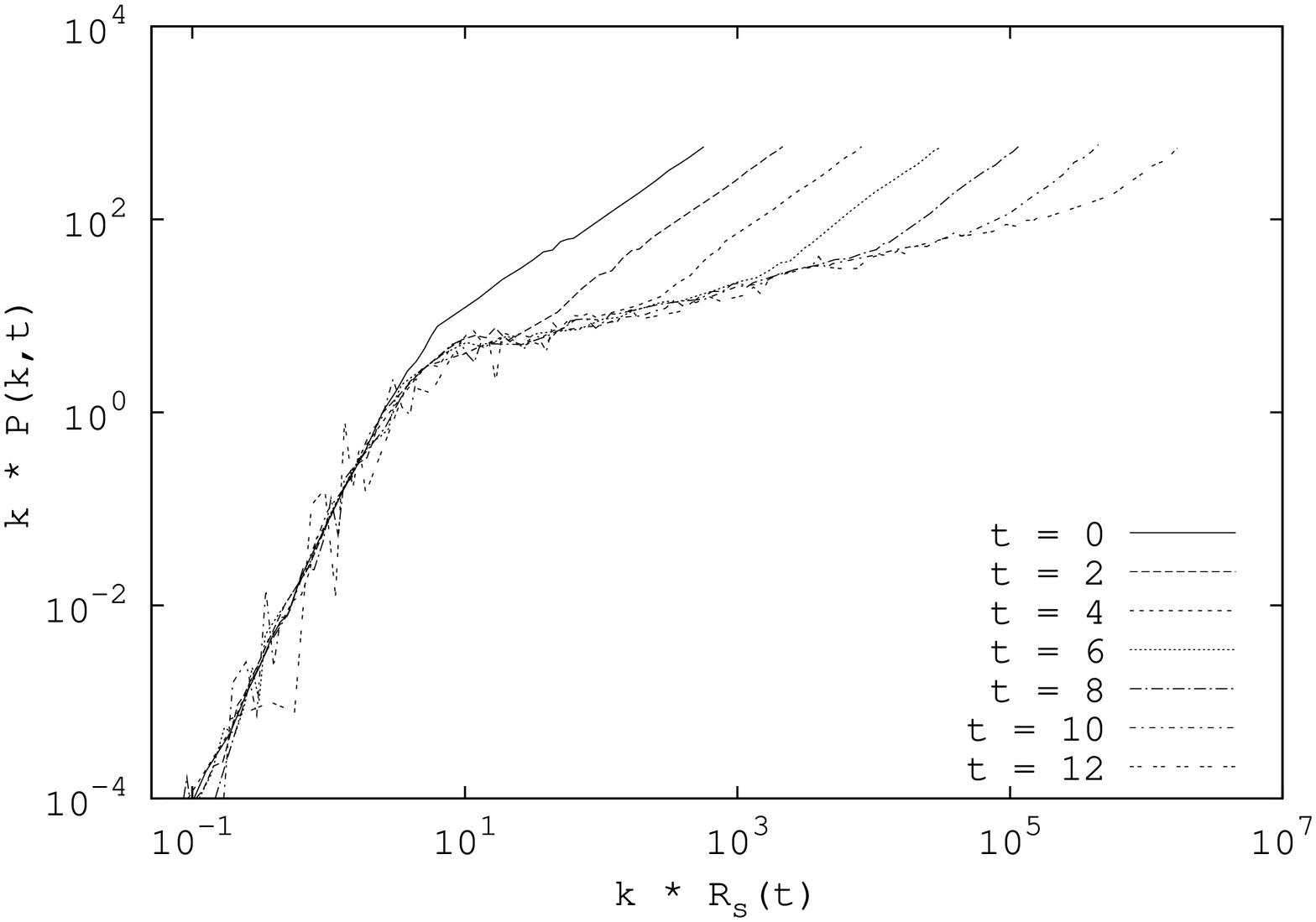}
\caption{The two point correlation function (upper panel) and power spectrum (lower panel) as 
a function of rescaled spatial variables, for the $\Gamma=1/\sqrt{6}$ model
and initial conditions with $n=2$ (i.e. EdS cosmology).  The ``time" variable
$t\equiv \log a$ (where $a=1$ initially). $L$ is the (periodic) box size. }
\label{CF_PS_k2_quintic}
 \end{figure}
\end{center}

Extending analyses in  \cite{ yano+gouda} 
and in \cite{agmjfs_pre2009} 
we have studied a range of values of initial PS  ($n=0, 2, 4$)
and models ($\Gamma=0, 1/\sqrt{6}, 1/\sqrt{2}$). The upper
panel in  Fig.~\ref{CF_PS_k2_quintic}  shows the measured reduced two point 
correlation function $\xi(x,t)$ as a function of $x/R_s(t)$,
while the lower panel shows the (dimensionless) 
$\Delta^2=kP(k)$ as a function of $kR_s(t)$, for the
case of an initial PS with $n=2$ and a model with
$ \Gamma=1/\sqrt{6}$. The superposition of the rescaled 
PS  at small $k$ in the lower panel shows
the validity of linear theory, while the superposition in the same
plot at larger $k$, and of $\xi(x)$ in the upper panel over a 
wide range,  shows the development of the corresponding 
self-similarity in the non-linear regime. We observed the
same behaviours in all cases, except for $n=4$. In the
latter case it turns out that, as expected, linear theory 
breaks down at small $k$, but nevertheless self-similarity 
is observed in the non-linear regime. 

In three dimensions such self-similar behaviour has been
observed numerically to apply in all cases which have
been simulated. 
Specifically in EdS universes 
it has been shown to apply in the range $-3<n\leq 1$
(see, e.g., \cite{efstathiou_88, colombi_etal_1996, jain+bertschinger_1998, 
smith}), while in the static universe limit it has been shown
to apply for $n=0$ and $n=2$ \citep{sl1, sl3}. 
As in all these 3D cases, a good power-law fit 
can be obtained (as in  Fig.~\ref{CF_PS_k2_quintic}) 
to the correlation function, $\xi(r) \sim r^{-\gamma}$,  and 
to the power  spectrum ( $k^dP(k) \sim k^{\gamma}$) in 
the range of non-linear self-similar clustering. 
The exponents $\gamma$ we measure in our 1D simulations,
which are reported below in further detail, show the same 
qualitative dependence on $n$ as in 3D:  in the expanding 
case $\gamma$ increases as $n$ does so (see, e.g., \cite{smith}),
while in the static case one observes a much smaller 
exponent which does not depend sensibly on $n$
\citep{sl1, sl3}. 

\section{Exponents of scale invariant clustering} 

Compared to 3D the power law behaviours we observe
in the 1D models extend  over a much larger range of scales,
simply because of the much greater dynamical range
accessible. In  Fig.~\ref{CF_PS_k2_quintic},
for example, such behaviour extends over roughly four decades
in scale, compared to a single decade in 3D studies 
(e.g. \cite{smith}). Such spatial resolution makes
it possible to robustly determine whether such behaviour
is indicative of scale invariant (i.e. fractal-type) properties
in the distributions, using tools such as box-counting methods
to measure the spectrum of  ``generalized dimensions"
(see, e.g. \cite{book}). In a recent study  of these 1D models with 
initial conditions like those we are studying here,  \cite{miller+rouet_2010a} 
have used such methods to analyze this clustering 
and have shown that, like in their previous studies
of slightly different initial conditions in \cite{miller_etal_2007}, 
they indeed find clear evidence for scaling behaviours indicative
of fractal clustering in this regime. 
We have performed much of the same analysis
on our simulations, and find results very consistent with
those reported in \cite{miller+rouet_2010a}:
robustly measured ``generalized dimensions" in the range
of scales where the power law behaviour in the
correlation function indicates scale invariance,
and in particular a value of the 
``correlation dimension"  $D_2 \approx 1-\gamma$ 
as expected.  

We show now that the exponent $\gamma$ measured 
from $\xi(x,t)$ can be very well accounted for, in the 
expanding models, using both the self-similarity of the 
evolution in time, and directly observable  ``stable clustering'' 
behaviour in the system at sufficiently small scales.  To do so 
let us consider the temporal evolution of the lower 
cut-off,  $x_{min}$ say, at which the self-similar 
scaling breaks down. Just as can be seen in 
Fig.~\ref{CF_PS_k2_quintic}  we observe in 
all our simulations that $\xi(x,t)$ breaks from
the power-law at this scale and become 
approximately flat, reaching a maximum value $\xi_{max}$. 
(In 3D simulations the effects of force smoothing make
it impossible to follow in this way the development of
self-similarity at small scales).
We observe to a very good approximation, in our simulations with
$\Gamma \neq 0$, that
\begin{equation}
\xi_{max}  \propto x_{min}^{-1} \sim \exp ({ -\epsilon \tau})
\label{xmin-ximax}
\end{equation}
where $\epsilon$ is a constant {\it which depends only 
on} $\Gamma$ (and not on $n$). Further we observe
that the upper cut-off to power-law behaviour in the 
correlation function, $x_{max}$, occurs (as in three
dimensions) at a fixed
amplitude in $\xi$ slightly larger than unity, i.e.,
roughly at the non-linearity scale [defined by 
$\xi (x_{NL}) =1$]. We therefore have, from self-similarity,
that  $x_{max} \propto R_s (t)$. It follows that 
the exponent $\gamma$, at sufficiently long times,
should be given by
\begin{equation}
\gamma = \frac{ \epsilon}{\epsilon  + \frac{2}{1+n} \lambda_+(\Gamma)}\,.
\label{gamma-epsilon}
\end{equation}

\begin{table*}
\begin{center}
\begin{tabular}{|c|c|c|c|c|c|c|}
\hline
intial PS	& $\Gamma=0$ (thy)	& $\Gamma=0$ (sim)		& $\Gamma=1/\sqrt{6}$ (thy)	& $\Gamma=1/\sqrt{6}$ (sim)	& $\Gamma=1/\sqrt{2}$ (thy)	& $\Gamma=1/\sqrt{2}$ (sim) \\
\hline
$n=0$ 		& $\gamma = 0$		& $\gamma = 0.18 \pm 0.03$	& $\gamma = 1/7$ 		& $\gamma = 0.14 \pm 0.02$	& $\gamma = 1/4$		& $\gamma = 0.25 \pm 0.02$ \\
$n=2$ 		& $\gamma = 0$		& $\gamma = 0.18 \pm 0.03$	& $\gamma = 1/3$		& $\gamma = 0.35 \pm 0.02$	& $\gamma = 1/2$		& $\gamma = 0.50 \pm 0.02$ \\
$n=4$ 		& $\gamma = 0$		& $\gamma = 0.15 \pm 0.05$	& $\gamma = 5/11$		& $\gamma = 0.43 \pm 0.01$	& $\gamma = 5/8$		& $\gamma = 0.62 \pm 0.01$ \\
\hline
\end{tabular}
\end{center}
\caption{Values of the exponent $\gamma(n,\Gamma)$ characterizing the power-law range in $\xi(x)$,
as measured in simulations (``sim") and as
predicted theoretically ("thy") by (\ref{gamma-sc-1D}), which gives $\gamma=(n+1)/(n+7)$ for $\Gamma=1/\sqrt{6}$,
and $\gamma=(n+1)/(n+4)$ for $\Gamma=1/\sqrt{2}$.
}
\label{table_result}
\end{table*}

To determine the exponent fully we need only determine how
$\epsilon$ in (\ref{xmin-ximax})  depends on $\Gamma$. 
The relation (\ref{xmin-ximax}) indicates the relevance of
 ``stable clustering" : if the correlation {\it up to this scale} $x_{min}$ arises from 
structures  which do not evolve (macroscopically) when considered
in spatial coordinates rescaled in proportion to $x_{min}$,  one obtains 
such a behaviour.  In three dimensions such a behaviour has been proposed 
long ago \citep{peebles_1974, davis+peebles_1977} as possibly valid 
{\it in the strongly non-linear regime}, and used to
obtain a prediction for the exponent $\gamma$ in an EdS 
universe. It amounts to supposing that if non-linear structures 
behave essentially as isolated systems (i.e. if the tidal 
forces exerted by all mass external to them are negligible),
they will be expected to virialize and remain stable in 
{\it physical} coordinates. 
In the 1D models
we are considering the equations of motion have not
been derived starting from physical coordinates, but
such ``stable clustering"  nevertheless has a clear 
meaning\footnote{We note that \cite{yano+gouda} give 
an incorrect generalization to 1D of the stable clustering 
hypothesis, assuming that stability will be attained in the 
3D physical coordinates.}.
Indeed in the 1D model the meaning 
of ``isolation" of a subsystem may be given more exactly 
than in 3D: if particles of a given subsystem 
do no cross particles outside the system, the corresponding 
system indeed evolves independently from the rest of 
the ``universe" (i.e. tidal gravitational forces vanish in 1D).
To see what evolution is then predicted, it is appropriate
to rewrite the equations of motion for such a finite 
isolated subsystem using the particle labelling in
which particles pass through one another.  
It is straightforward to show that the equation of motion
for particles belonging to such a subsystem
may then be written (in the units we have adopted
in which $g=1/2n_0$)
\begin{equation}
\frac{d^2 {x}_i}{d\tau^2} +
\Gamma \frac{d { x}_i}{d\tau} =  g(N^>(i) - N^<(i)) +  x_i
\end{equation}
where $x_i$ is the position of the particle $i$ {\it with respect
to the centre of mass of the system}, and $N^>(i)$ ($N^<(i)$) 
is the number of particles (in the subsystem) on the right (left) 
of the particle $i$. Unsurprisingly this is just the equation of motion 
for a {\it finite} 1D self-gravitating system in an 
infinite space, with an additional damping term arising from the 
expansion and a term from the subtracted mean density.
We can define the conserved energy $E$ associated with 
the two terms on the right hand side, and have then
\begin{equation}
\frac{d {E}}{d\tau} = -  2\Gamma K
\label{energy-equation}
\end{equation}
where $K$ is the total kinetic energy, and $E=K+U_{grav}+U_{bg}$
with $U_{grav}=\frac{1}{2} \sum_{i,j} g |x_i - x_j|$ and 
$U_{bg}=\sum_i x_i^2/2$. Now,  if such an isolated  subsystem 
is significantly {\it overdense} (i.e. its mean density $n_s$ is much
greater than $n_0$), the time scale associated with the dynamics 
of  the first term on the right hand side $\tau_{grav} \sim \sqrt{n_0/n_s}$ 
is much shorter than that associated with the other two terms,
and the term $U_{bg}$ is negligible compared to $U_{grav}$.
We expect this to allow the use of an adiabatic approximation
to (\ref{energy-equation}), in which $E$ and $K$ are replaced
by their values averaged on this shorter time scale, 
$\langle E \rangle$ and $\langle K \rangle$.
On these time scales, however, we expect the corresponding
virial  relation to apply, i.e., $\langle 2K \rangle =\langle U _{grav}\rangle$.
It follows that $\langle E \rangle=3\langle K \rangle$,
and then, using Eq.~(\ref{energy-equation}),  that 
\begin{equation}
\langle K \rangle  \propto \exp \left(-\frac{2\Gamma \tau}{3} \right) \propto \langle U_{grav} \rangle \,.
\end{equation}
Given that $U_{grav}$ is linearly proportional to the size of the subsystem,
the size of such a virialized subsystem will be expected to scale  
in the same way. Thus, if the behaviour
in (\ref{xmin-ximax}) is indeed due to such stable virialized 
structures, $\epsilon=2 \Gamma/3$.
Substituting this relation in  (\ref{gamma-epsilon}) gives 
\begin{equation}
\gamma = \frac{ 2\Gamma(n+1)}{ \Gamma (2n-1)  +
3 \sqrt{\Gamma^2 +4}} \,.
\label{gamma-sc-1D}
\end{equation}
We note that, using the above arguments, a 
straightforward generalization of the usual 3D stable 
clustering prediction (for an EdS cosmology) to the 
class of cosmological backgrounds corresponding 
to any (constant) $\Gamma$ may be obtained.
The result is\footnote{Virialization in three dimensions leads 
to $ \langle E \rangle = - \langle K \rangle$, so that
$ \langle U_{grav} \rangle \propto \exp \left(2\Gamma \tau \right)$.
It follows that  $x_{min} \propto \exp (-2 \Gamma \tau)$, and then, 
with  $\xi_{max} \propto x_{min}^{-3}$ and the 3D expression 
for $R_s(t)$, the result follows.}
\begin{equation}
\gamma_3 = \frac{ 6\Gamma(n+3)}{ \Gamma (2n+5)  +
\sqrt{\Gamma^2 +4}} \,
\label{gamma-sc-3D}
\end{equation}
which indeed coincides with the usual EdS 
prediction,  $\gamma=3(3+n)/(5+n)$, for $\Gamma=1/\sqrt{6}$.

Shown in Table \ref{table_result} is a comparison of the
prediction (\ref{gamma-sc-1D}) with the corresponding
exponents measured in our simulations.
The agreement is 
excellent for the  expanding cases, and indeed in these cases the value
$\epsilon$ measured directly agrees well with the 
predicted $\epsilon=2\Gamma/3$~\footnote{For 
the case $\Gamma=1/\sqrt{6}$ our measured 
exponents agree with those reported 
by \cite{yano+gouda}. Our stable clustering result 
$\epsilon=2\Gamma/3$ explains also the exponent
measured by \cite{aurell_etal_2003} directly 
for an isolated structure.}.
In the static case the prediction leads to the badly
defined result $\gamma=0$, while we observe a small
exponent  $\gamma_0 \approx 0.16$ which does not appear
to depend on $n$. We observe in this case that, rather
than (\ref{xmin-ximax}), we have an $x_{min}$ which remains
approximately constant (i.e.  the expected $\Gamma=0$
limit), but a $\xi_{max}$ which grows. 
Clearly the stable clustering hypothesis is indeed only
an approximation, to which corrections (due to the fact 
that structures are not, of course, strictly isolated) become
most important, as 
one would expect, when we go to the limit 
$\Gamma \rightarrow 0$ in which virialized structures
no longer decrease in size. 

\section{ Halos and virialization in the scale invariant regime } 
Let us now consider the observed clustering in terms of ``halos",
in analogy to how this is done in cosmological simulations.
We thus use a friend-of-friend algorithm, with a variable
linking length $\Lambda$ (which we define in units of 
the initial lattice spacing $\ell$). In 1D this is equivalent to 
separating the system into structures separated by voids of size
greater than or equal to $\Lambda$. In the range of
scale-invariant non-linear clustering, i.e., for 
$x_{min}/\ell < \Lambda < x_{NL}/\ell$, we observe
that the algorithm selects out structures with a broad 
range of sizes from our simulations, peaked around a 
value which scales with $\Lambda$. 
However, the fact that the distribution of sizes is quite large, while the 
range of scale invariance
in any simulation is of course quite limited, make
it difficult (even in 1D) to use such
halos as a tool for characterizing this regime.
We thus analyse here the case of an initial PS 
with $n=4$ in which this range is greatest ---
a little more than four orders of magnitude 
in scale between  
$x_{min}/\ell \approx 10^{-2.5}$ and 
$x_{NL}/\ell \approx 10^{1.5}$ in the most
evolved configuration. 

\begin{figure*}
\begin{center}
 \includegraphics[angle=-90, width=2\columnwidth]{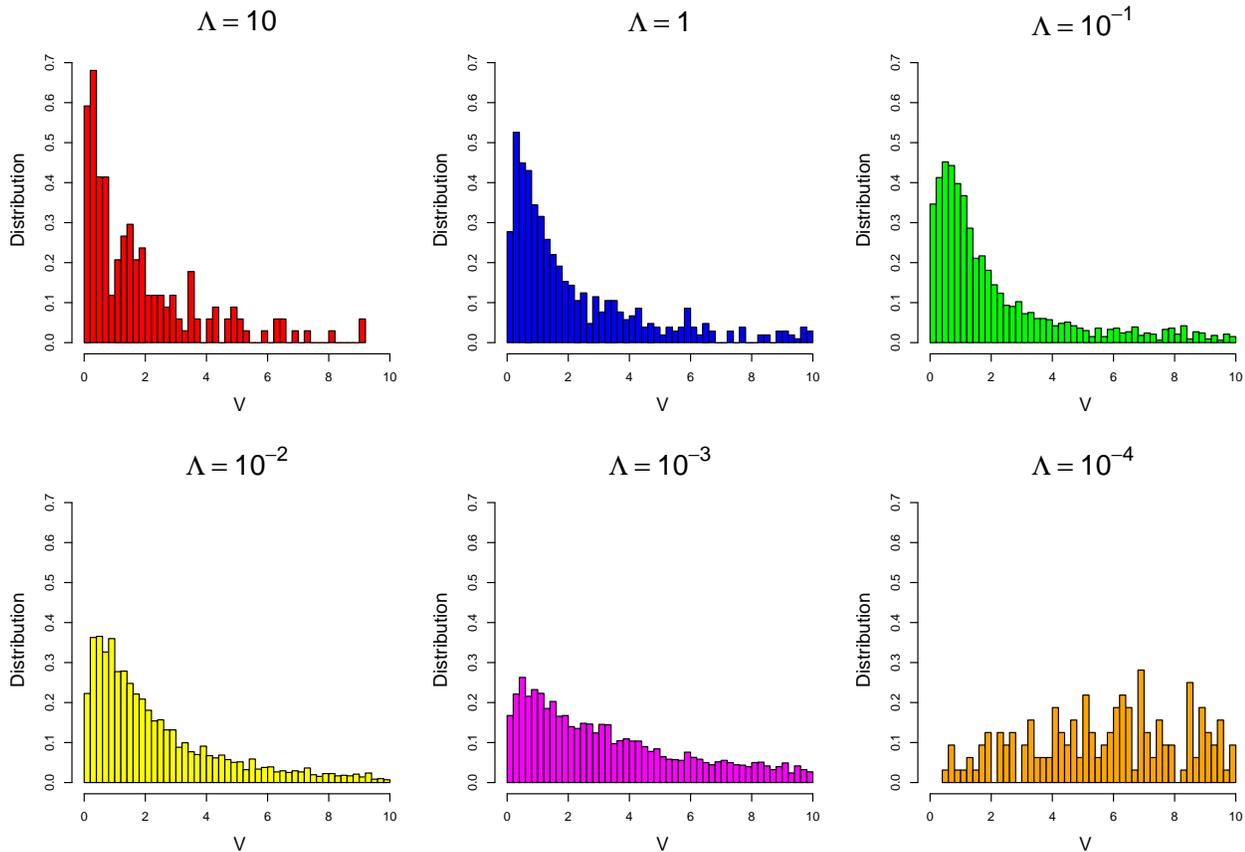}
\caption{Normalized distribution of the virial ratios of halos selected with the 
indicated values of $\Lambda$ (linking length in units of lattice spacing), for the 
case $n=4$ and $\Gamma=1/\sqrt{6}$ at $t=\ln a=22$. 
}
\label{virial_ratio}
\end{center}
\end{figure*}

Shown in Fig.~\ref{virial_ratio} is the 
measured distribution of virial ratio as
a function of $\Lambda$ in this simulation,
i.e. the distribution of the values of the
ratio $2K/U_{grav}$ measured directly in
each halo, taking the velocities with respect
to the centre of mass of the corresponding 
halo. We observe an approximately stable
distribution {\it peaked about unity} in the 
range of $\Lambda$ where the structures 
selected out fall mostly in the length scales 
of scale invariance, while at the smallest  
scales no such tendency is observed.
On the contrary the virial ratio tends to
be very large, which can be interpreted
as due to the fact that the algorithm is
selecting structures inside the scale
$x_{min}$: if the distribution at this
scale is constituted of approximately 
smooth virialized structures of size
$x_{min}$,  parts of these structures
will have a high virial ratio (because
$K$ scales more slowly with size 
than $U_{grav}$ in a smooth
structure).

We have studied quantitatively the stability of these 
measured distributions in the range of scale invariant clustering 
using a nonparametric Kolmogorov-Smirnov (K-S) test.
The K-S statistic quantifies a distance between the empirical 
distribution functions of the different samples, and the null distribution 
is calculated under the null hypothesis that the samples are 
drawn from the same distribution.  Calculating the corresponding 
$p$-values of this test, we find that the null hypothesis is indeed
not rejected in the range of scale-invariance, while,
on the other hand, it is clearly rejected outside this range.

Our interpretation of these results is that one can effectively 
decompose the distribution into a collection of structures
which are, approximately, virialized. This is true provided 
these structures are defined at any scale in the range 
of scale-invariant clustering.

\section{Discussion}

{\it In three dimensions} the stable clustering 
hypothesis has been used widely as a reference point,  and 
indeed much used phenomenological models such as the
formalism proposed by \cite{peacock} 
incorporate it into the modeling of the non-linear power spectrum 
obtained from numerical simulations. However, in parallel,
results of major simulations of power law initial conditions
(e.g. \cite{efstathiou_88, smith}) led to the conclusion that, 
although the measured exponents in the correlation function 
showed a behaviour roughly consistent with its predictions (albeit
with some significant deviations \citep{smith}), there was no
evidence for the ``clustering hierarchy" which
Peebles had argued would be associated 
with it (see, e.g. \cite{peebles}).  \cite{efstathiou_88}, notably, 
explicitly excluded such a ``fractal" description and 
found evidence instead  for the validity of a description in terms 
of ``smooth non-linear clumps". These latter are the precursors 
of the ``halos" of halo models,  which have become the standard 
phenomenological description of the matter distribution in cold dark matter
cosmologies (for a review, see, e.g. \cite{halo}). The matter distribution
is then approximated as a collection of isolated (and thus virialized)
spherical structures with smooth density profiles. Further, on the basis
of extensive numerical study following that of \cite{navarro2}, the latter 
are widely believed  to be characterized well by ``universal"  exponents 
 (i.e. independent of initial conditions and cosmology). 

\begin{figure*}
\begin{center}
  \includegraphics[angle=-90, width=2\columnwidth]{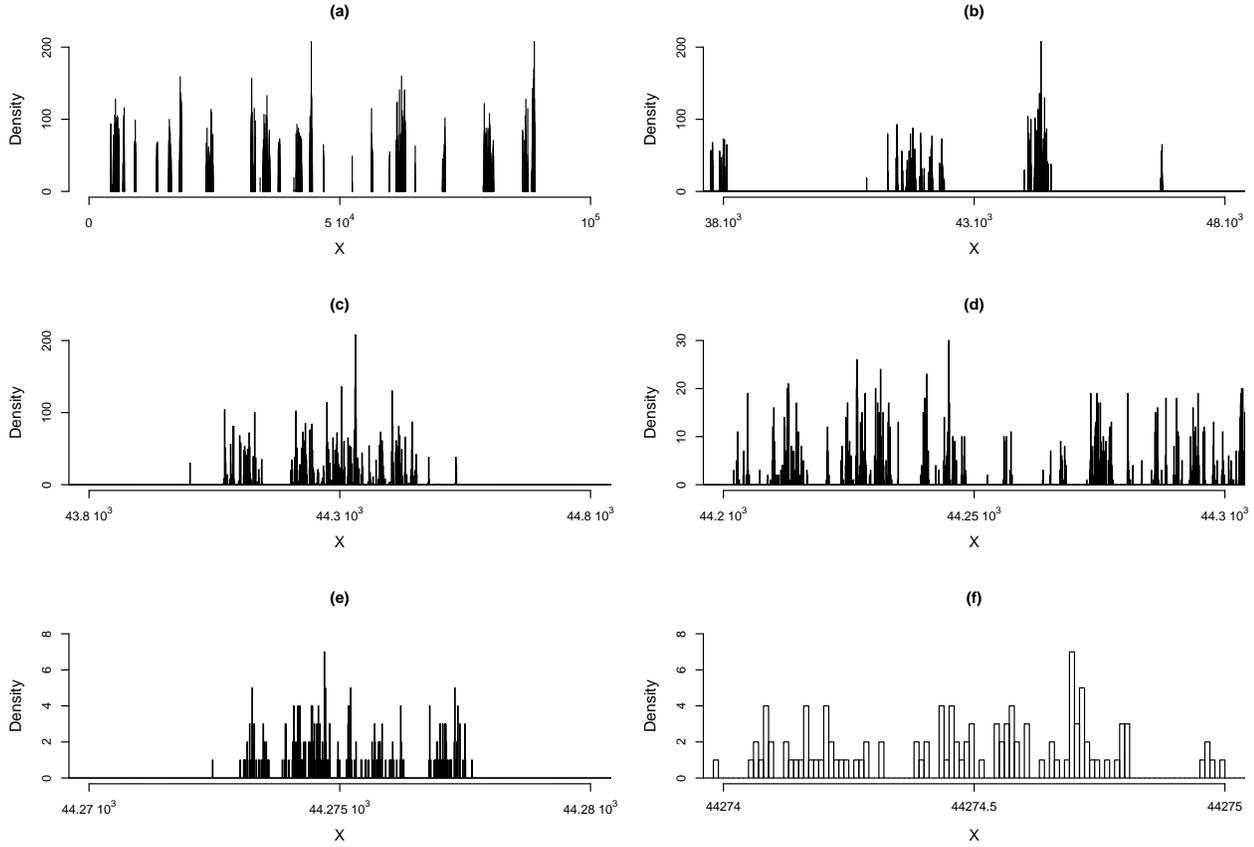}
\caption{ Density field obtained at $t=\log a=14$ starting from an initial condition
with $n=2$, in the ``EdS" model with $\Gamma=1/\sqrt{6}$. The first panel shows 
the whole box (i.e. length units are such that  $L=N=10^5$), while each subsequent
panel shows a spatial ``zoom"  on a region of size {\it one tenth} that shown 
in the previous panel. The resolution has been increased (i.e. the bin size has been 
decreased) in the last three plots  in order to reveal the clumpy nature of the 
distribution at these scales
}
\label{Zoom_density_fractal}
\end{center}
 \end{figure*}
 
{\it In one dimension } the strongly non-linear regime is
truly scale-invariant (for power law initial conditions) in
a range which grows monotonically in time. The associated 
distribution of matter is {\it intrinsically} lumpy or grainy
down to the lower cut-off  scale $x_{min}$:  indeed 
the very meaning of such {\it scale invariance}  is that there
are no characteristic scales available to define
 smoothness.   To illustrate this
we show in Fig. ~\ref{Zoom_density_fractal} 
the spatial density at various levels of
detail in a typical evolved configuration. Such 
distributions are most naturally described with the
instruments of fractal (or, more generally, multi-fractal) 
analysis developed for this purpose in condensed matter 
and statistical physics (see e.g. \cite{book}). 
As we have just illustrated in the previous section, halo type 
descriptions may, of course,  also be employed to describe them. 
The halos so defined are, however, very different to those described
by 3D halo models:  these 1D halos are not smooth. 
Further have they no intrinsic size themselves, but are 
defined only with reference to an arbitrary chosen 
scale.  The study of the virial ratios we have presented 
indicates, however, that such halos can be considered as 
entities with a dynamical relevance, as they show a clear 
tendency to have a virial ratio of order unity (which is the 
behaviour of an isolated structure). The clustering in
the non-linear regime can thus be considered as
a concrete realization of the qualitative picture of 
a  ``clustering hierarchy" originally envisaged by 
Peebles (e.g. \cite{peebles}) as resulting from stable
clustering. The stable clustering hypothesis we have described 
above, however, is actually subtly different from the original one: 
we assumed only that stable clustering applies below
the scale $x_{min}$ marking the lower cut-off to the 
scale invariance, and not necessarily to  the strongly non-linear
regime as a whole. Thus we assumed only  that stable clustering 
applies at an ultraviolet scale {\it fixed by the  resolution of the 
simulation} (or, physically, by the scale at which
the very first structures form).  The ``statistical virialization" we have observed 
using the halo analysis, on the other hand,  applies at scales 
above $x_{min}$ and across the range of the scale 
invariant  clustering.

There are clearly two possible conclusions one can draw
from this analysis: 

\begin{itemize}
\item A. These 1D models produce 
non-linear clustering which is qualitatively different in its nature
to that in 3D, or
\item B. The spatial resolution in 3D simulations up to now
has been too limited to reveal the nature of clustering in
cold dark matter cosmologies, which is correctly reflected
(qualitatively) in the 1D simulations.
\end{itemize}

We believe that, despite the impressive computational size 
and sophistication of 3D cosmological simulations, 
conclusion B may well be the correct one. The very largest modern studies 
in a cosmological volume access roughly two decades in scale
in the non-linear regime\footnote{ For example, the ``millennium" 
simulation \citep{springel_05} of the $\Lambda CDM$ cosmology,
has,  at redshift zero,  $\xi=1$ at about $2$h$^{-1}$Mpc,  an 
initial mean interparticle $\ell  \approx 250$h$^{-1}$
kpc and a force smoothing $\epsilon=5$h$^{-1}$ kpc. It is usually
supposed that it is the latter which sets the lower bound on
resolution, but we note that it has been shown by several
studies  (see \cite{melott_etal_1997, romeo_2008,
mjbmtb_2009}) that, at the very least, precision may be compromised
below the scale $\ell$.}  while reference studies in the literature of 
power law initial conditions in EdS cosmology \citep{efstathiou_88, smith} 
measure the crucial power-law  behaviour in the correlation function (or the PS) 
over at  most one decade.  If we were to perform our 1D simulations at  
comparable resolution to large cosmological simulations
like \cite{smith}, we would certainly have great difficulty in establishing 
the scale-invariant nature of the strongly non-linear clustering
arising from power law initial conditions. Although halos 
defined exactly as in three dimensions might look clumpy,  
an approximately smooth profile could be determined for 
them if they were averaged (as they can be in three dimensions when 
spherical symmetry is assumed).  Higher resolution 3D simulations of 
smaller regions have shown over the last decade that there is in fact 
much more substructure (``subhalos") inside halos than was originally anticipated
 (see, e.g., \cite{moore_etal_1999, diemand_etal_2005, goerdt_etal_2007}),
and have even more recently described several levels of such
substructure (``subhalos of subhalos", see e.g.  \cite{diemand_etal_nature_2008,
springel_etal_aquarius_2008, stadel_etal_2009}). 
\cite{diemand_etal_nature_2008} even use the term ``fractal"
to describe (qualitatively) the real space structures, while 
\cite{zemp_etal_2009}  describe the structure of halos
in phase space as ``intrinsically grainy". We note that 
other authors 
(see, e.g., \cite{valageas_1999, gaite_2007}) have 
previously  argued for similar conclusions on the
basis of analyses of 3D simulations.

Let us consider nevertheless one possible consideration
in favour of (the more conservative) conclusion A. In the 
expanding (i.e. damped)  1D models, the stable
clustering prediction (\ref{gamma-sc-1D}) fits the 
measured exponents extremely well. 
Early 3D studies  for EdS cosmologies (e.g. \cite{efstathiou_88}) 
measured exponents roughly consistent with the stable clustering 
prediction, but later studies (e.g. \cite{smith}) have found significant 
disagreement. This disagreement is attributed to physical 
mechanisms which cause the fundamental assumption of
stability to be violated --- by the evident fact that {\it there
are}  interactions between ``halos", which can even lead
to their merging into single structures. We have noted that
in one dimension  tidal forces vanish, and structures 
can interact only when they actually physically
cross one another. While merging may occur, 
it may be that it is a less efficient process than in three
dimensions. Thus the excellent agreement in the 1D
models 
compared to EdS may perhaps be attributed to the 
fact that these models probably represent poorly
the role of such physical effects. The essential question, 
however, is not whether these effects 
play a role and can
lead to deviations from stable clustering, but whether such
effects can {\it qualitatively} change the nature of clustering,
destroying scale invariance by smoothing out the distribution
{\it on a scale related to the upper cut-off} to scale invariance. 
Our study of the case $\Gamma=0$ suggests that the 
answer is negative. The prediction of stable clustering
does not work in this case, and like in three dimensions,
one obtains a small value of the exponent which does
not sensibly depend on $n$. The physical reasons why the 
exponent is close to, but different to, the stable clustering 
prediction are a priori the ones just cited.
The analysis of Miller et al.
of this case, which we have rechecked and confirm, finds
nevertheless that the distribution is scale invariant.
Further, as we have mentioned, the lower cut-off 
$x_{min}$ remains constant as in the stable clustering 
hypothesis, of order the initial lattice spacing (and
unrelated to the upper cut-off).

These results on 1D models suggest directions for
3D investigations which might establish definitively
the correctness of conclusion B.
We note, for example, that the 1D models lead one to
expect that the exponents derived phenomenologically to 
characterize the highly non-linear density field inside 
smoothed halos (i.e. the ``inner slope" of halos) should 
be closely related to the exponent $\gamma$ determined 
from the correlation function.  Indeed --- in the approximation
of a simple fractal behavior in the strongly non-linear regime,  
which the spectrum of multi-fractal exponents measured in
 \cite{miller+rouet_2010a} suggests should be quite good --- 
the mean density about the centre of such halos will decrease just
as about any random point, i.e., with the {\it same} 
exponent $\gamma$.  Despite the contradiction with
the widely claimed ``universality" of such exponents in 
halos profiles, such a hypothesis cannot currently
be ruled out,  as the determination of such exponents 
is  beset by numerical difficulties (arising again 
from the limited resolution of numerical simulations).
In a study of halo profiles obtained from power law 
initial conditions \cite{knollmann_etal_2008} show
explicitly that the results for the halo exponents 
depend  greatly on what numerical fitting procedure is 
adopted. While one procedure gives ``universality"  
(i.e. exponents independent of $n$), a different one 
favors clearly steepening inner profiles for larger $n$. 
Indeed we note that the numerical values for
the inner slopes obtained by \cite{knollmann_etal_2008} 
are, for the larger $n$ investigated, in quite good 
agreement with the exponent predicted by 
stable clustering.
 
Our results here are for power law initial conditions, but
we could equally use the model to study both initial conditions
and an appropriately modified time-dependent damping rate
mimicking the $\Lambda$CDM model in one dimension. The 
exact scale invariance in the strongly non-linear regime
would certainly be broken in this case, and the various
characteristic scales introduced will be imprinted 
on the clustering. There would thus not be a single
correlation exponent, but a slowly varying one. One would 
certainly expect, however, the qualitative nature of the 
clustering to be unchanged,
just as in three dimensions clustering appears to be
qualitatively the same in ``scale-free" cold dark matter
cosmologies and $\Lambda$CDM. 
Further,  our considerations here are strictly relevant only
to dissipationless {\it cold} dark matter simulations. If the initial 
conditions are ``warm" or ``hot", or if other non-gravitational 
interactions are turned on, the associated physical effects will 
tend to smooth out the matter distribution up to some
scale (and thus destroy the scale invariance up to
this scale). 
Nevertheless, if the conclusion B is correct
even for this idealized case, it is likely to have very
important observational implications relevant to
testing standard cosmological models. Intrinsically
clumpy or grainy halos lead, for example, to very different
 predictions for dark matter annihilation (see, 
 e.g. \cite{goerdt_etal_2007, diemand_etal_nature_2008, afshordi_etal_2010})). 
Concerning the compatibility of such a distribution of cold
dark matter with observations of the distribution of visible
matter at sub-galactic scales --- which would be expected to
lead to an amplified  ``missing satellite" problem  --- we note 
that recent studies such as \cite{wadepuhl+springel_2010}
show that solid conclusions in this respect will be reached
only on the basis of a much improved understanding of
the processes involved in star formation. 
At larger scales the possible link to the striking power-law
behavior which characterizes galaxy correlations 
over several decades in scale
(see, e.g., \cite{peebles_1974, sylos_etal_1998, masjedi_etal_2006}) 
--- which was the motivation for original work using stable clustering
to explain such behaviour from power law initial conditions \citep{peebles_1974} and is naturally interpreted
as indicative of  underlying scale invariance in the matter distribution
(see, e.g. \cite{sylos_etal_1998, gaite_2007, antal_etal_2009})  --- is intriguing, 
and will be discussed elsewhere.

We acknowledge useful discussions with Andrea Gabrielli, Jean-Michel
Levy,  Francesco Sylos Labini and Tirawut Worrakitpoonpon. We thank 
the anonymous referee for several very useful suggestions.
\bibliographystyle{mn2e}


\end{document}